\documentclass[12pt]{iopart}

\usepackage{times}
\usepackage{graphicx}

\newcommand{\GQPC}{G_{\rm QPC}}

\newcommand{\Rser}{R_{\rm ser}}

\begin{document}

\title[Temperature dependent deviations from ideal quantization of plateau 
conductances... ]
{Temperature dependent deviations from
ideal quantization of plateau conductances in GaAs quantum point
contacts}  

\author{A.~E.~Hansen$^{1,2}$, A.~Kristensen$^{1,3}$, and H.~Bruus$^{1,3}$}
\address{$^{1}$Niels Bohr Institute, Univ.\ Copenhagen, Universitetsparken 5,
DK-2100 Copenhagen\\
$^{2}$Institute of Physics, University of Basel,
Klingelbergstrasse 82,
CH-4056 Basel\\
$^{3}$Mikroelektronik Centret, Technical University of Denmark,
{\O}rsteds Plads, DK-2800 Lyngby}
\begin{abstract}

We present detailed experimental studies of the temperature
dependence of the plateau conductance
of GaAs quantum point contacts
in the temperature range from 0.3~K to 10~K. Due to a strong
lateral confinement produced by a shallow-etching technique we are
able to observe the following unexpected feature: a linear temperature
dependence of the measured mid-plateau conductance.
We discuss an interpretation
in terms of a temperature dependent, intrinsic series
resistance, due to non-ballistic effects in the 2D-1D transition
region. These results have been
reproduced in several samples from different
GaAs/GaAlAs-heterostructures and observed in different
experimental set-ups.

\end{abstract}

\section{Introduction}

More than a decade after the first observation of quantized
conductance, the quantum point contact (QPC) remains a central
research topic in mesoscopic physics. QPCs are often employed as
basic components in mesoscopic experiments, and more recently they
are playing a central role in studies of spin-related physics such
as spin filtering \cite{Thomas}, quantum entanglement
\cite{Yamamoto}, the Kondo effect \cite{Marcus2002}, and
single-photon generation \cite{SAW}. These applications
necessitate a detailed understanding of the QPC, and this is the
motivation for our work.

In this paper we present experimental observations of a novel
temperature effect in GaAs QPCs: a linear temperature dependence
of the measured mid-plateau conductance. As in previous work
\cite{Kristensen2000} our new experiments rely on a shallow-etch
technique in the fabrication of the QPC, which yields particularly
strong lateral confinement. This enables studies of the
temperature dependence of the quantized QPC conductance up to 10~K.

\begin{figure}[b]
\centerline{
\includegraphics[height=42mm]{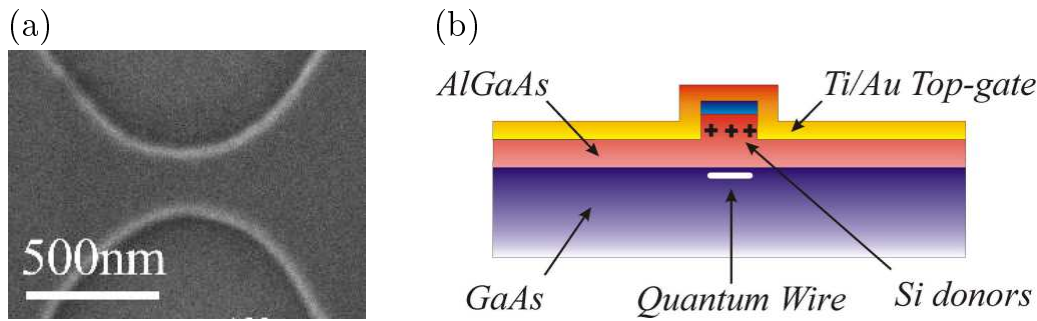}}
\caption{\label{fig:device} (a) SEM picture of the etched quantum point
contact, before deposition of the top gate. 
(b) A cross section of the device.}
\end{figure}

\section{Sample fabrication}

The QPCs were fabricated on 2-dimensional electron gases
(2DEGs) defined in modulation doped GaAs/GaAlAs
heterostructures. Typical 2DEG mobilities and densities were
70-100~m$^2$/Vs and 2$\times$10$^{15}$ m$^{-2}$, respectively.
The narrow QPC constrictions were defined by
shallow wet etching as described in Ref.~\cite{Kristensen2000}.
The phenomena discussed here were observed in three different
experimental set-ups and on several samples. We present data
from one particular sample,
shown in Fig.~\ref{fig:device}a. The electron density in the shallow
etched QPC-constriction (approximately 200~nm wide and 200~nm
long) is controlled by a 10~$\mu$m wide Ti/Au topgate electrode, with
an applied voltage denoted $V_{\rm QPC}$, covering that region.
The samples have an energy spacing of the two first transverse
subbands of 5-10~meV. The
samples are situated on a 20~$\mu$m wide and 100~$\mu$m long mesa,
with several ohmic contacts.

\begin{figure}[ht]
\centerline{\includegraphics[height=90mm]{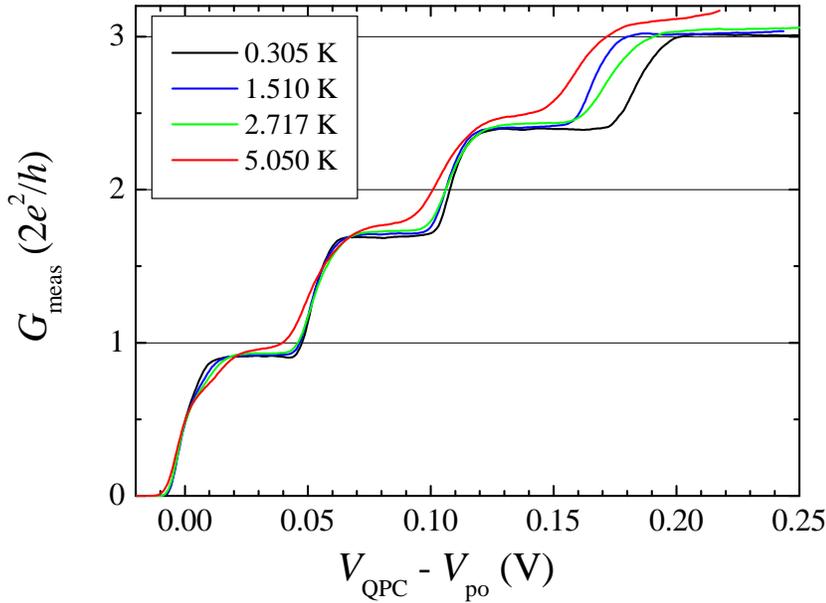}}
\caption{\label{fig:rawdata} Measured conductance, $G_{\rm meas}$,
versus the QPC gate voltage, $V_{\rm QPC}$ at the temperatures 
given in the legend.
}
\end{figure}

\section{Measurements}

The sample was mounted in a pumped $^3$He cryostat, enabling
measurements in the temperature range from 0.3 to 10~K. The
differential conductance, $dI/dV_{\rm sd}$ was measured in a 2-point
voltage controlled setup, using a standard lock-in technique at
117~Hz. Here $V_{\rm sd}$ is the source-drain voltage bias and $I$
the corresponding current.

In Fig.~\ref{fig:rawdata} we show raw data plots of the
gate-characteristics, i.e.\ the measured conductance, $G_{\rm
meas}$, versus gate voltage, $V_{\rm QPC} - V_{\rm po}$, taken at
different temperatures. The pinch-off voltage, $V_{\rm po}$,
is defined as $G_{\rm meas}(V_{\rm po}) = e^2/h$.
For the raw data, the conductance plateaus
do not coincide with the expected integer quantization. This is,
in part, due to an inevitable series resistance.

\begin{figure}
\centerline{\includegraphics[height=80mm]{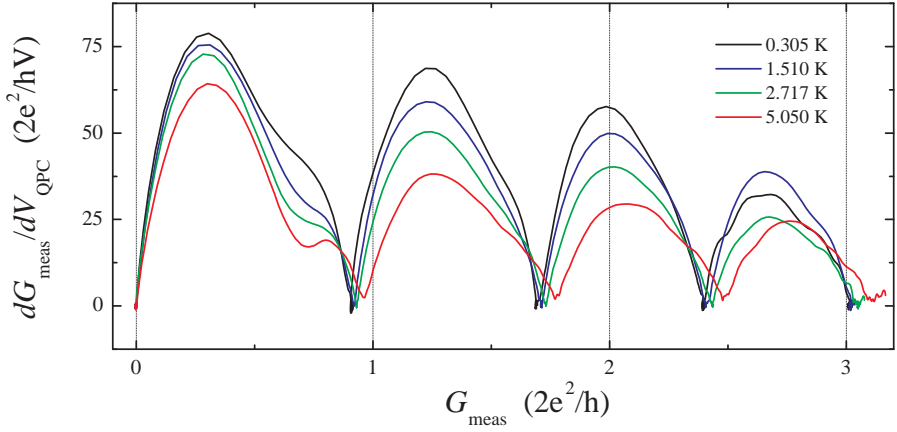}}
\caption{\label{fig:transcond} Transconductance,
$dG_{\rm meas}/dV_{\rm QPC}$, versus conductance, $G_{\rm meas}$,
used to identify point on the conductance plateaus with minimum
slope. }
\end{figure}

We want to study the plateau conductance, which we define by the
conductance value at the point of minimal slope of the gate
characteristics. To extract these values, we plot the
transconductance, $dG_{\rm meas}/dV_{\rm QPC}$, versus
conductance, $G_{\rm meas}$, as shown in Fig.~\ref{fig:transcond}.
We note that on this plot the sharp minima, identifying the plateau
conductance on the four plateaus, exhibit a remarkable increase
with temperature. This temperature dependence is the central issue
of our work.



\section{Data analysis}

In the data analysis we use the plateau conductance values to
extract information about the series resistance. However, the
observed temperature dependence of the plateau conductances
complicates this analysis. In general, the measured resistance,
$R_{\rm meas} = 1/G_{\rm meas}$, can be written as a
combination
of the resistance of the QPC, $1/G_{\rm QPC}$, and the series
resistance, $R_{\rm ser}$:
\begin{equation}\label{Rmeas}
  R_{\rm meas} = \frac{1}{G_{\rm QPC}} + R_{\rm ser}.
\end{equation}
A crucial point is to relate the plateau conductance $\GQPC(N)$ of
the $N$'th plateau to the conductance quantum, $2e^2/h$. As seen
in Fig.~\ref{fig:rawdata} and Fig.~\ref{fig:transcond}, 
the plateaus are well defined, which
can justify the most simple assumption,
\begin{equation}\label{GQPCN}
  \GQPC(N) = \frac{2e^2}{h}\: N .
\end{equation}
In this case, which is analyzed in Sec.~\ref{sec:GN}, the observed
temperature dependence of the plateau conductances is attributed
solely to the series resistance.

The temperature dependence could also be interpreted in terms of a
temperature dependent coefficient, $\alpha(T)$, in front of the
conductance quantum
\begin{equation}\label{GQPCNT}
  \GQPC(N,T) = \alpha(T) \: \frac{2e^2}{h}\: N .
\end{equation}
This assumption is analyzed in Sec.~\ref{sec:GNT}.

\subsection{Temperature and gate voltage dependent $R_{\rm ser}$ and
  constant coefficient $\alpha$}
\label{sec:GN}

We begin with the most simple assumption, namely Eq.~(\ref{GQPCN})
stating that the QPC conductance $\GQPC(N)$ is given by $N$ times
the conductance quantum. This assumption is justified by the high
quality of the observed conductance plateaus. In terms of
Eq.~(\ref{GQPCNT}) this corresponds to fixing the coefficient
$\alpha(T) = 1$, so the only free parameter in Eq.~(\ref{Rmeas}) is
the series resistance $R_{ser}$.

\begin{figure}
\centerline{
\includegraphics[height=85mm]{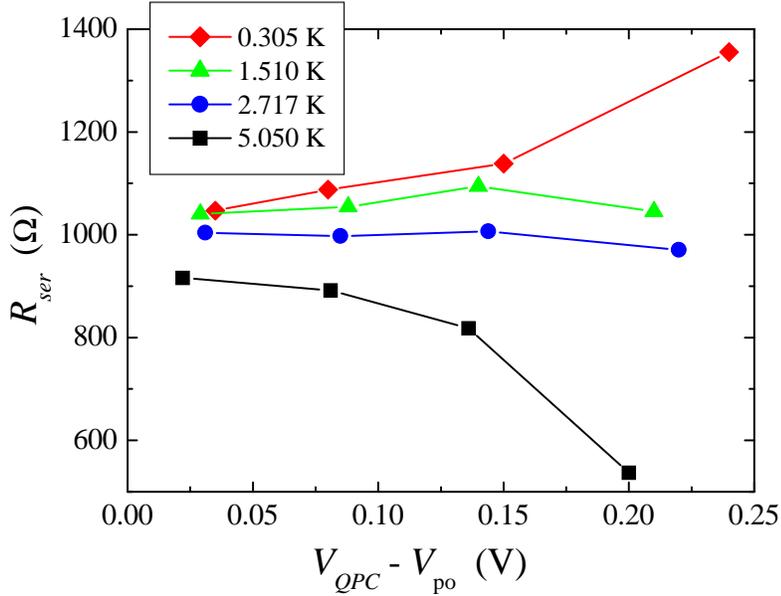}}
\caption{\label{fig:GN} Temperature and gate-voltage dependent
series resistance $R_{\rm ser}(T,V_{\rm QPC})$ assuming that the
plateau conductance is given by Eq.~(\ref{GQPCN}).}
\end{figure}

For each mid-plateau conductance extracted from
Fig.~\ref{fig:transcond}, it is now possible to calculate a value
of $R_{ser}$. For a given plateau at a given temperature we assign
to $R_{ser}$ the value of $V_{\rm QPC}$ displaced by $V_{\rm po}$
as in Fig.~\ref{fig:rawdata}. Thus we obtain a temperature and
gate voltage dependent series resistance $\Rser(T,V_{\rm QPC})$
shown in Fig.~\ref{fig:GN}. The series resistance is seen to
depend strongly on temperature and gate voltage.

We keep in mind that the measured conductance include
approximately 100$\times$20~$\mu$m of 2DEG, in addition to the QPC
constriction. So the question naturally arises if the series
resistance in Fig.~\ref{fig:GN} could be a bulk 2DEG resistance.
For temperatures smaller than the maximum temperature of 
10~K used here, the resistance
of the unpatterned 2DEG is basically constant. The strong
temperature dependence of $R_{\rm ser}$ seen in Fig.~\ref{fig:GN}
(a change of more than a factor of $2$ at the last plateau), makes
it very implausible that the series resistance is solely due to
the resistance of the 2DEG.

The same conclusion can be drawn from the gate voltage dependence.
The gate controlling the electron density in the QPC also covers
approximately 10$\times$20~$\mu$m  of the 2DEG. Therefore the series
resistance of the 2DEG will change with gate voltage by a fraction
of the ungated square resistance, $\sim$ 50 $\Omega$. But the
resistance of the 2DEG will be monotonically decreasing with
increasing gate voltage. This is contradiction with the data points
in Fig.~\ref{fig:GN} for the lowest temperatures.

We conclude that series resistance extracted from the measured
conductance plateaus, shows a remarkable dependence on temperature
and gate voltage. It cannot be attributed to the resistance of the
bulk 2DEG, in which the quantum point contact is situated. We will
discuss other interpretations of the extracted series resistance in
Sec.~\ref{discuss}.

\subsection{Temperature dependent resistance $R_{\rm ser}$ and
  coefficient $\alpha$}
\label{sec:GNT}

\begin{figure}
\centerline{
\includegraphics[width=0.6\textwidth]{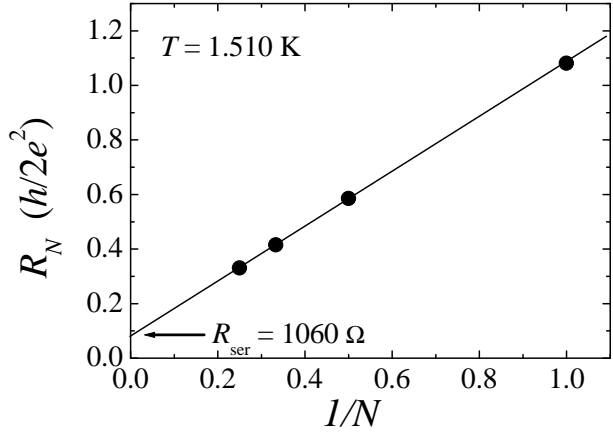}}
\caption{\label{fig:linreg}  Linear fit to plateau resistance $R_{N}$
versus inverse plateau number $1/N$ at fixed temperature, $T$,
resulting in determination of the series resistance, $\Rser(T)$,
and the conductance coefficient $\alpha(T)$.}
\end{figure}

\begin{figure}
\begin{center}
\includegraphics[width=0.48\textwidth]{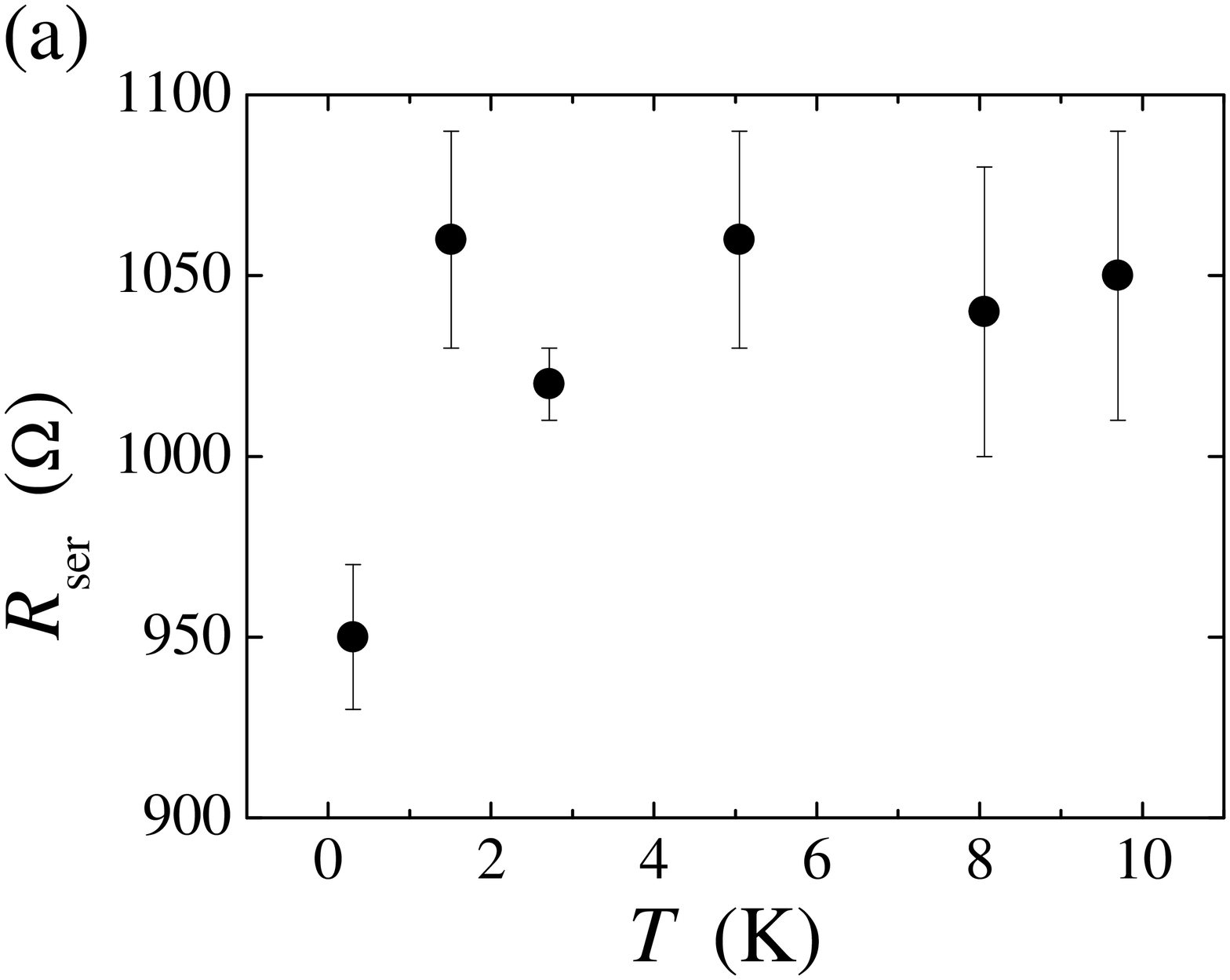}
\includegraphics[width=0.48\textwidth]{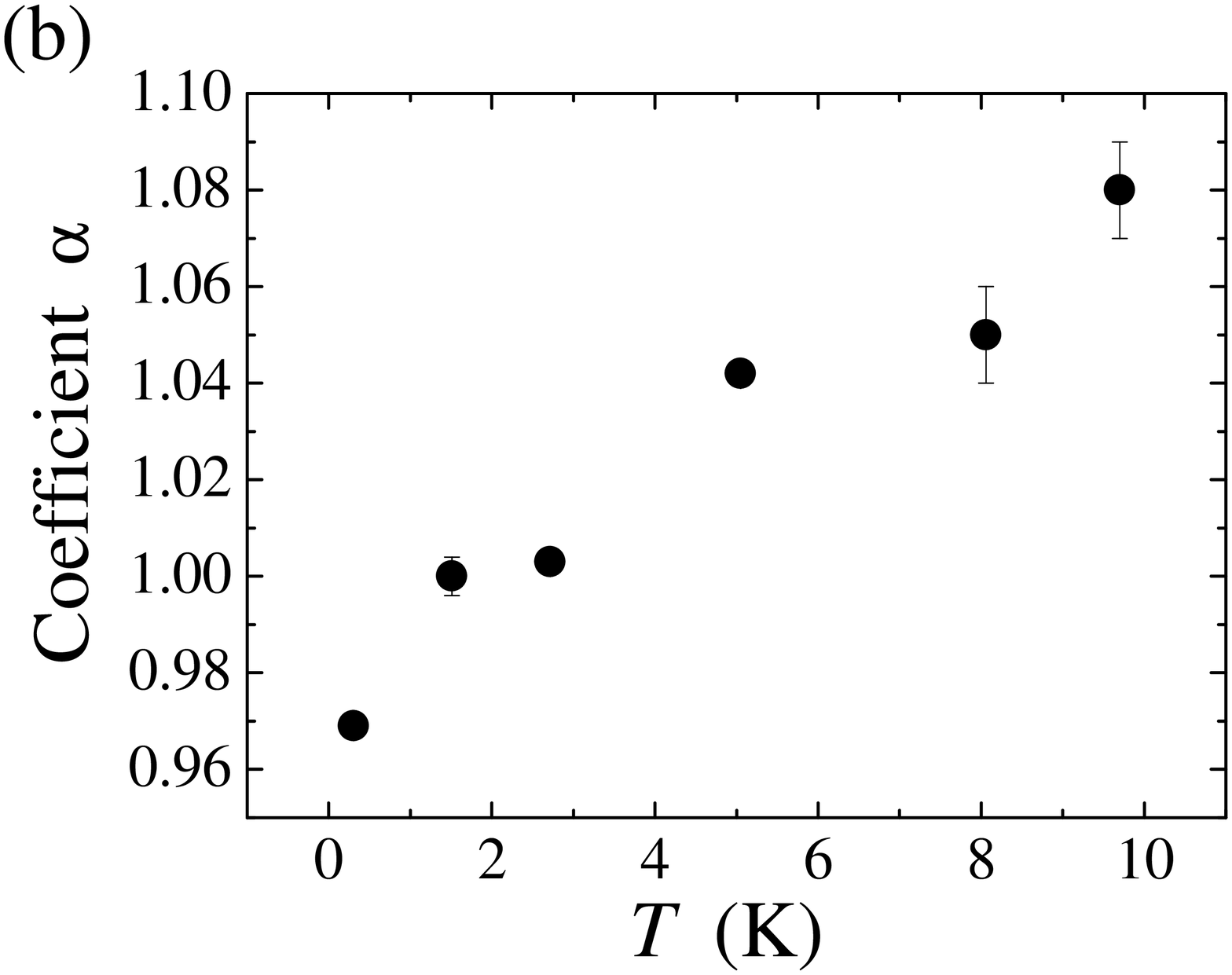}
\end{center}
\caption{\label{fig:RserAlpha}  Data analysis based on
Eq.~(\ref{GQPCNT}) and linear fits like the one shown in
Fig.~\ref{fig:linreg}. (a) Plot of
$\Rser(T)$ versus $T$. (b) Plot of $\alpha(T)$ versus $T$.}
\end{figure}

We now allow $\alpha$ in Eq.~(\ref{GQPCNT}) to depend on
temperature, and assume $\Rser$ to be independent of gate voltage
but varying with temperature. We plot the resistance value of the
$N$'th conductance plateau $R(N)$, as function of $1/N$ in
Fig.~\ref{fig:linreg}. With Eq.~(\ref{Rmeas}) and~(\ref{GQPCNT})
this allows us to extract $\Rser(T)$ and $\alpha(T)$, shown in
Figs.~\ref{fig:RserAlpha}a and \ref{fig:RserAlpha}b, respectively.
The series resistance basically remains constant as function of
temperature. However, $\alpha$ increases linearly as temperature
is raised from $0.3$ to $10$~K.

A temperature dependent plateau conductance of quantum wires has
earlier been reported, e.g., in Ref.~\cite{Tarucha}, and
interpreted in the context of Luttinger liquid physics combined
with disorder in the wires. Here a similar interpretation is not
possible. The Luttinger liquid theory applies for extended 1D
systems, whereas our system  only has a length of $\sim$ 200~nm,
which is a few Fermi wavelengths and similar to the width.
Additionally, the effect of such electron-electron correlations
should be a suppression of the conductance below the quantized
value. Here we observe a conductance plateau that at one
temperature is clearly below the quantized value and at another
clearly above. This effect is larger than the error bar on the
calibration of the resistance measurement setup. We note that a
temperature dependent plateau conductance has also been reported
for a different quantum wire system \cite{Yacoby96}. Here the
conductance was only reduced below the quantized value ($\alpha <
1$), a result that could be interpreted in terms of scattering
across a 2D-1D interface \cite{Vadim97,Picciotto00}.

We conclude that the temperature dependence of the height of the
conductance plateaus, can be described by a temperature dependent
overall rescaling of the conductance, and a temperature and gate
voltage independent series resistance.

\section{Discussion}
\label{discuss}

We have presented measurements of quantized conductance in a
shallow-etched GaAs QPC in the temperature range from 0.3~K to
10~K. The four well-defined conductance steps show on close
inspection a temperature dependent deviation from the expected
quantization value. The deviation cannot be accounted for by
describing our system as a simple series connection of a 2DEG and
an ideal quantum point contact. We have observed this effect in
several different devices, with both 2-point and  4-point
measurement schemes.

We can assign correct values to the measured conductance plateaus
by allowing an overall conductance rescaling factor $\alpha$ to
vary (a few percent) with temperature. The small length of our
device and the fact that we can measure $\alpha$ greater than
unity depending on temperature, rules out Luttinger liquid
corrections as the cause of our findings. We can alternative
describe the deviations from the quantized conductance values by
an effective series resistance $\Rser$, which depends strongly on
gate voltage and  temperature. We underline that the parameter
dependence of $\Rser$ is incompatible with that of a pure bulk
2DEG resistance.

The above discussion has rejected intrinsic 1D or 2D effects as
the origin of the observed deviations from conductance
quantization. This suggests that non-ballistic effects in the
1D-2D contact region are the cause instead. The high quality of
the conductance pleateaus makes it implausible that reflection
takes place in the QPC itself. Reflections in the contact region
can of course only decrease the plateau conductance below its
quantized value. The non-trivial temperature and gate voltage
dependence of the resulting effective series resistance
(Fig.~\ref{fig:GN}), which results from this interpretation,
points to an intricate effective potential landscape in the
contact region, which deserves further experimental and
theoretical investigation. We remark that the nature of the 1D-2D
contact region is very different from that of the quantum wires in
Refs.~\cite{Yacoby96,Picciotto00}, so in our case is not possible
to take over straightforwardly the interpretations presented in
Refs.~\cite{Vadim97,Picciotto00}.

\section*{Acknowledgements}
We thank C.B. S{\o}rensen for growing the semiconductor
structures, and P.E. Lindelof for hosting the experiments. This
work was partly supported by the Danish Technical Research Council
(grant no.~9701490) and by the Danish Natural Science Research
Council (grants no.~9502937, 9600548 and 9601677). The III-V
materials used in this investigation were made at the III-V
NANOLAB, operated jointly by the Microelectronics Centre,
Technical University of Denmark, and the Niels Bohr Institute,
University of Copenhagen.


\section*{References}
\begin{thebibliography}{9}

\bibitem{Thomas} K.J. Thomas et al.,
  Phys. Rev. Lett. {\bf 77}, 135 (1996)

\bibitem{Yamamoto} R.C. Liu et al.,
  Nature {\bf 391}, 263 (1998)

\bibitem{Marcus2002} S.M. Cronenwett et al.,
  Phys. Rev. Lett. {\bf 88}, 226805 (2002)

\bibitem{SAW} EU-project SAW-PHOTON,
  http://ntserv.fys.ku.dk/sawphoton/

\bibitem{Kristensen2000} A. Kristensen et al.,
  Phys. Rev. B {\bf 62}, 10950 (2000)

\bibitem{Tarucha} S.\ Tarucha et al., Solid State Comm.\ {\bf 94}, 413 (1995)

\bibitem{Yacoby96} A. Yacoby et al.,
  Phys. Rev. Lett. 77, 4612 (2000)

\bibitem{Vadim97} A. Yu. Alekseev and V. V. Cheianov,
  Phys. Rev. B 57, R6834 (1998)

\bibitem{Picciotto00} R. de Picciotto et al.,
  Phys. Rev. Lett. 85, 1730 (2000)

\end{thebibliography}
\end{document}